\documentclass{PoS}

\title{Color Confinement and Supersymmetric Properties of Hadron Physics 
from Light-Front Holography}

\ShortTitle{Supersymmetric Properties of  QCD and Light-Front Holography}

\author{\speaker{Stanley J. Brodsky}\thanks{
Supported in part by the Department of Energy, contract No. DE-AC02-76SF00515.
.}\\
        Stanford Linear Accelerator Center, Stanford University, Stanford, CA, 94309\\
        E-mail: \email{sjbth@slac.stanford.edu}}

\abstract{I review  applications of  superconformal algebra. light-front holography, and an extended form of conformal symmetry to hadron spectroscopy and dynamics.  QCD is not supersymmetrical in the traditional sense -- the QCD Lagrangian is based on quark and gluonic fields -- not squarks nor gluinos. 
However, its hadronic eigensolutions conform to a representation of superconformal algebra, reflecting the underlying conformal symmetry of chiral QCD.   
The  eigensolutions of superconformal algebra provide a unified Regge spectroscopy of meson, baryon, and tetraquarks of the same parity and twist as equal-mass members of the same 4-plet representation with a universal Regge slope.  
The pion $q \bar q$ eigenstate is composite but yet has zero mass for $m_q=0.$   
Light-Front Holography also predicts  the form of the nonperturbative QCD running coupling:  $\alpha_s(Q^2) \propto \exp{-{Q^2/4 \kappa^2}}$,  in agreement with the effective charge  determined from measurements of the Bjorken sum rule.  One also obtains viable predictions for tests of hadron dynamics such as spacelike and timelike hadronic form factors, structure functions, distribution amplitudes, and transverse momentum distributions. 
The combined approach of light-front holography and superconformal algebra also provides insight into the origin of the QCD mass scale and color confinement.  
A key tool is the dAFF principle which shows how a mass scale can appear in the Hamiltonian and the equations of motion while retaining the conformal symmetry of the action.  When one applies the dAFF procedure to chiral QCD, a mass scale $\kappa$ appears which determines universal Regge slopes, hadron masses in the absence of the Higgs coupling.  The result is an extended conformal symmetry which has a conformally invariant action even though an underlying mass scale  appears in the Hamiltonian.  Although conformal symmetry is strongly broken by the heavy quark mass, the
supersymmetric mechanism, which transforms 
mesons to baryons (and
baryons to tetraquarks), still holds and gives remarkable mass degeneracies across the spectrum of light, heavy-light and double-heavy hadrons.
 }

\FullConference{Light Cone 2019 - QCD on the light cone: from hadrons to heavy ions - LC2019\\
		16-20 September 2019\\
		Ecole Polytechnique, Palaiseau, France}

\begin{document}

\section{Introduction}

\begin{figure}[h]
\includegraphics[width=18pc]{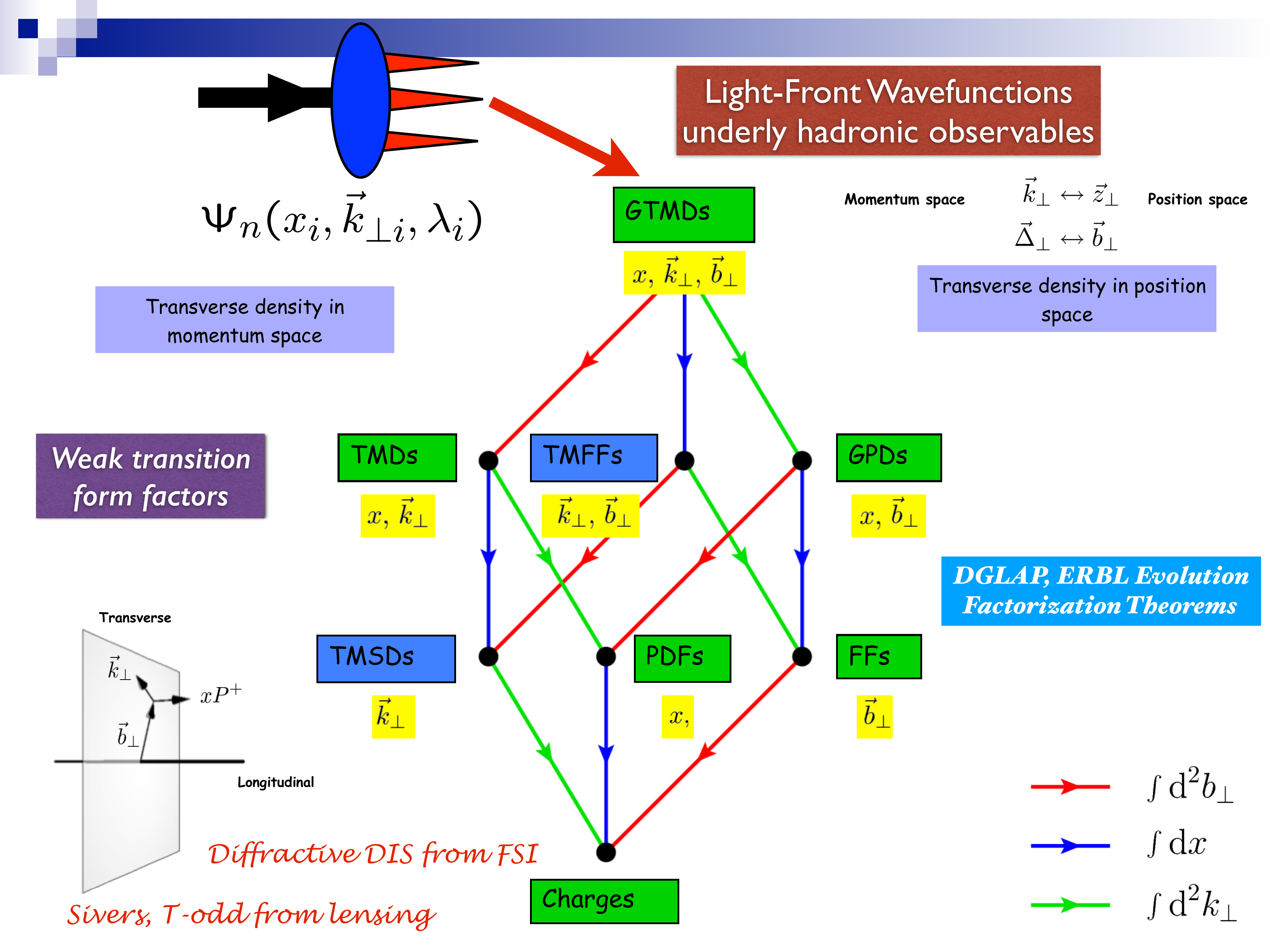}\hspace{2pc}%
\begin{minipage}[b]{18pc}\caption{\label{LorcePasquini}.   Light-Front wavefunctions encode hadron structure and underlie hadron observables such as the Drell-Yan-West Formula for elastic and inelastic  form factors, structure functions, generalized parton distribution, etc. 
Observables with complex phases,  such as diffractive deep inelastic scattering $e p \to e' p' X$ and  the Sivers  pseudo-T-odd spin  correlation  $\vec S_p \cdot \vec q \times \vec p_q$, shadowing and antishadowing of nuclear structure function,  incorporate the Wilson lines which involve final and/or initial state interactions, as well as the LFWFs. 
Adopted from an illustration by B. Pasquini and C. Lorc\'e~\cite{Lorce:2012jy,Lorce:2011ni}.}
\end{minipage}
\end{figure}

One of the most important theoretical tool in high energy physics is Dirac's light-front time  $\tau =x^+ = t +z/c $,  the time along the light-front~\cite{Dirac:1949cp}, a concept which allows all of the tools and insights of 
Schr\"odinger's nonrelativistic quantum mechanics and the Hamiltonian formalism to be applied to relativistic physics~\cite{Brodsky:1997de}.  
When one takes a photograph, the object is observed at a fixed light-front (LF) time.  Similarly,  Compton  scattering and deep-inelastic lepton scattering on a hadron
are measurements of hadron structure at fixed LF time. Unlike ordinary ``instant time" $t$,
physics at fixed $\tau$ is Poincar\'e invariant; i.e.,  independent of the observer's Lorentz frame.   LF time  $\tau$ reduces to ordinary time $t$ in the nonrelativistic limit $c\to \infty.$
The LF wavefunctions of hadrons $\Psi^H_n(x_i, \vec k_{\perp i,} \lambda_i) = <\Psi_H|n>$ are the Fock state projections of the eigensolution of the QCD
LF Hamiltonian $H_{QCD}|\Psi_H >= M^2_H \Psi_H >$.  They encode the underlying structure of bound states in quantum field theory and underlie virtually every observable in hadron physics.  See Fig.~\ref{LorcePasquini}.  
Hadronic LFWFs can also be measured directly by the Ashery method~\cite{Ashery:2005wa}, the coherent diffractive dissociation of high energy hadrons into jets~\cite{Bertsch:1981py,Frankfurt:2000jm}.

Hadronic LFWFs are defined at fixed $\tau =-x^+ = t+z/c$;
they are thus off-shell in the total $P^- =P^0 -P^z$, instead of energy $P^0$.  The LF 3-momenta $P^+ = P^0+P^z$ and $\vec P_\perp$ are conserved. Thus LFWFs are also off-shell in
${\cal M}^2=P^+ P^- -P^2_\perp =  [\sum_i k^\mu_i]^2 = \sum_i {k^2_\perp+ m^2\over x}\vert_i,$   
the invariant mass squared of the constituents in the $n$-particle Fock state. The LFWFs are thus functions of the invariant mass squared of the constituents in the Fock state. For a two-particle Fock state:
${\cal M}^2 = {k^2_\perp+ m^2 \over x(1-x)}$.   Thus the constituent transverse momenta  $k^2_{\perp i} $ do appear alone as a separate factor in the LFWF; the transverse momenta are always always coupled to the longitudinal LF momentum fractions $x_i$: 
the light-front analog of rotational invariance.  Only positive $ k^+_i  =k^0_i+k^z_i \ge 0$ and $0 \le x_i ={k^+-\over P^+}  \le 1$  appear in the LF Fock states,   with
$\sum_i x_i = 1. $ 
In addition, $J^z = \sum_i { L^z_i+S^z_i}$,  as well as $P^+ =\sum_i k^+_i$ and $\vec P_\perp= \sum_i \vec k_{\perp i}$ are conserved at every vertex, essential covariant kinematical constraints. The LF spins are quantized in the $z$ direction, in contrast to ordinary Wick helicity.  Only one power of orbital angular momentum $L^z$ can appear at a vertex in a renormalizable theory.   This leads to  new rigorous selection rules  for  the spin dependence of scattering amplitudes~\cite{Chiu:2017ycx}.
One also can demonstrate an important property of quantum gravity applied to hadrons:  the anomalous gravitomagnetic moment of every LF Fock state of a hadron vanishes at $Q^2=0$~\cite{Lowdon:2017idv}.

The LFWFs of bound states are off-shell in $P^- \ne \sum k^-_i$, but they tend to be maximal at minimal off-shellness; i.e., minimal invariant mass ${\cal M}^2 $. In fact, in the holographic LFWFs where color is confined, the LFWFs of hadrons have fast Gaussian fall-off in invariant mass~\cite{deTeramond:2008ht}. This feature of minimal off-shellness of the LFWFs also underlie intrinsic heavy quark Fock states~\cite{Brodsky:1980pb}: the LFWFs have maximal support when all of the constituents
have the same rapidity $y_i$; i.e., $x_i \propto \sqrt{m^2_i + k^2_{\perp i}}$. Thus, in contradiction to the usual gluon splitting  $g \to Q \bar Q$ mechanism, the heavy quarks have the highest momentum fractions $x_i$.

There are many other important physics properties which become explicit using the light-front formalism~\cite{Brodsky:1997de}, such as 
 the ``color transparency " of hard exclusive reactions~\cite{Brodsky:1988xz}, 
 the ``hidden color" of nuclear eigenstates~\cite{Brodsky:1983vf}, the $q(x) \ne \bar q(x)$  asymmetry of non-valence sea-quark distributions~\cite{Brodsky:1996hc}, and the local  two-photon``seagull"  (``$J=0$ fixed-pole") contributions in the Compton amplitude from LF instantaneous quark exchange interactions\cite{Brodsky:2008qu}. The DGLAP evolution of  structure functions and the ERBL evolution of distribution amplitudes are most easily derived using LF  Hamiltonian theory~\cite{Lepage:1980fj}.  
In addition, LF perturbation theory  has  the special property of  ``history":  One does not  need to compute the numerator of any perturbative LF amplitude more than once, since only the LF denominator changes from one computation to another.

\section{Color Confinement, Extended Conformal Covariance,  and the Origin of the Hadronic Mass Scale}

If one sets the quark masses to zero in the Lagrangian of quantum chromodynamics (QCD), no hadronic mass scale is evident. 
Thus a fundamental question for QCD is the origin of the mass of the proton and other hadrons.   It is often stated that  the mass scale $\Lambda_{\overline{MS}}$ of the renormalized perturbative theory generates the nonperturbative QCD mass scale; however, this `dimensional transmutation" solution is problematic 
since the perturbative scale is renormalization-scheme dependent, whereas hadron masses cannot depend on a theoretical convention.   
It is often argued that the QCD mass scale reflects the presence of quark and gluon condensates in the QCD vacuum state.  
However, such condensates lead to a cosmological constant a factor of $10^{42}$ larger than measured.  In fact, up to LF zero modes, such as the Higgs background field~\cite{Srivastava:2002mw},
nontrivial vacuum structure does not appear in QCD if one defines the vacuum state as the eigenstate of lowest invariant mass of the QCD light-front (LF) Hamiltonian.  
In fact, in Dirac's boost invariant``front form" ~\cite{Dirac:1949cp},   where the time variable is the time $x^+ = t+z/c$ along the light-front, the light-front vacuum $|0>_{LF}$ is both causal and frame-independent;  one thus has $<0_{LF} |T^{\mu \nu} | 0_{LF}> = 0$~\cite{Lowdon:2017idv} and zero cosmological constant~\cite{Brodsky:2009zd,Brodsky:2012ku}. 

The LF zero modes correspond to a constant scalar background with zero energy and three-momentum.   In the case of the Higgs theory, the traditional Higgs vacuum expectation value (VEV) is replaced by a ``zero mode", in the LF theory, analogous to a classical 
Stark or Zeeman field~\cite{Srivastava:2002mw}.   This Higgs LF zero mode (the LF analog of the  Higgs VEV )  gives mass to fermions via their Yukawa couplings.

In general,  one can reproduce the LF Hamiltonian results from  covariant Feynman-Lagrangian theory and the Bethe-Salpeter formalism for bound states 
by first performing the $k^-$ integration of the 4-dimensional $\int d^4 k$ loop integrals and picking up the pole contributions. However, there is an important  exception: in the case of vacuum loops,  there are  ``circle at infinity " contributions~\cite{Mannheim:2019lss}  to the Feynman loop integration which do  not appear unambiguously from the vacuum loop diagrams of LF Hamiltonian perturbation theory.  For example, in scalar field theories such as $g\phi^4$ theory, these non-pole contributions cause vacuum amplitudes  to have ``zero mode"  $k^\mu=0$  contributions which are not  given by a light-front Hamiltonian Fock space analysis.  There is thus an interesting distinction between the  vacuum structure of conventional Feynman-Lagrangian theory and LF Hamiltonian theory\cite{Collins:2018aqt,Mannheim:2019lss}.

It is conventional to measure hadron masses in MeV units; however, QCD has no knowledge of units such as electron-volts.  Thus QCD at $m_q=0$ can at best only predict ratios of masses such as $m_\rho/m_p$ and other dimensionless quantities. 
The  work of de Alfaro, Fubini, and Furlan  (dAFF)~\cite{deAlfaro:1976je}  provides a novel solution for the origin of the hadron mass scale in QCD.  They showed that one can introduce a nonzero mass scale $\kappa$ into the Hamiltonian of a conformal theory without affecting the conformal invariance of the action.   
The essential step of this ``extended conformal invariance" is to add to the Hamiltonian $H$ a term proportional to the dilation operator and/or the special conformal operator.  In the case of one-dimensional quantum mechanics, the resulting Hamiltonian acquires a confining harmonic oscillator potential $\kappa^4 x^2$; however, after a redefinition of the time variable, the action remains conformal.    The new time variable has finite range, consistent with the finite LF time between the constituents in a confining theory. The mass  scale $\kappa$ is not determined and serves as a holding parameter-- only ratios of eigen-masses are predicted.

The same principle  of ``extended conformal invariance" can be applied to relativistic quantum field theory using light-front (LF) quantization~\cite{Brodsky:1997de}. 
De T\'eramond, Dosch, and I~\cite{Brodsky:2013ar}
have shown that a mass gap and a fundamental color confinement scale also appear when one extends the dAFF procedure to light-front (LF) Hamiltonian theory  in physical 3+1 spacetime. 

\begin{figure}[h]
\includegraphics[width=18pc]{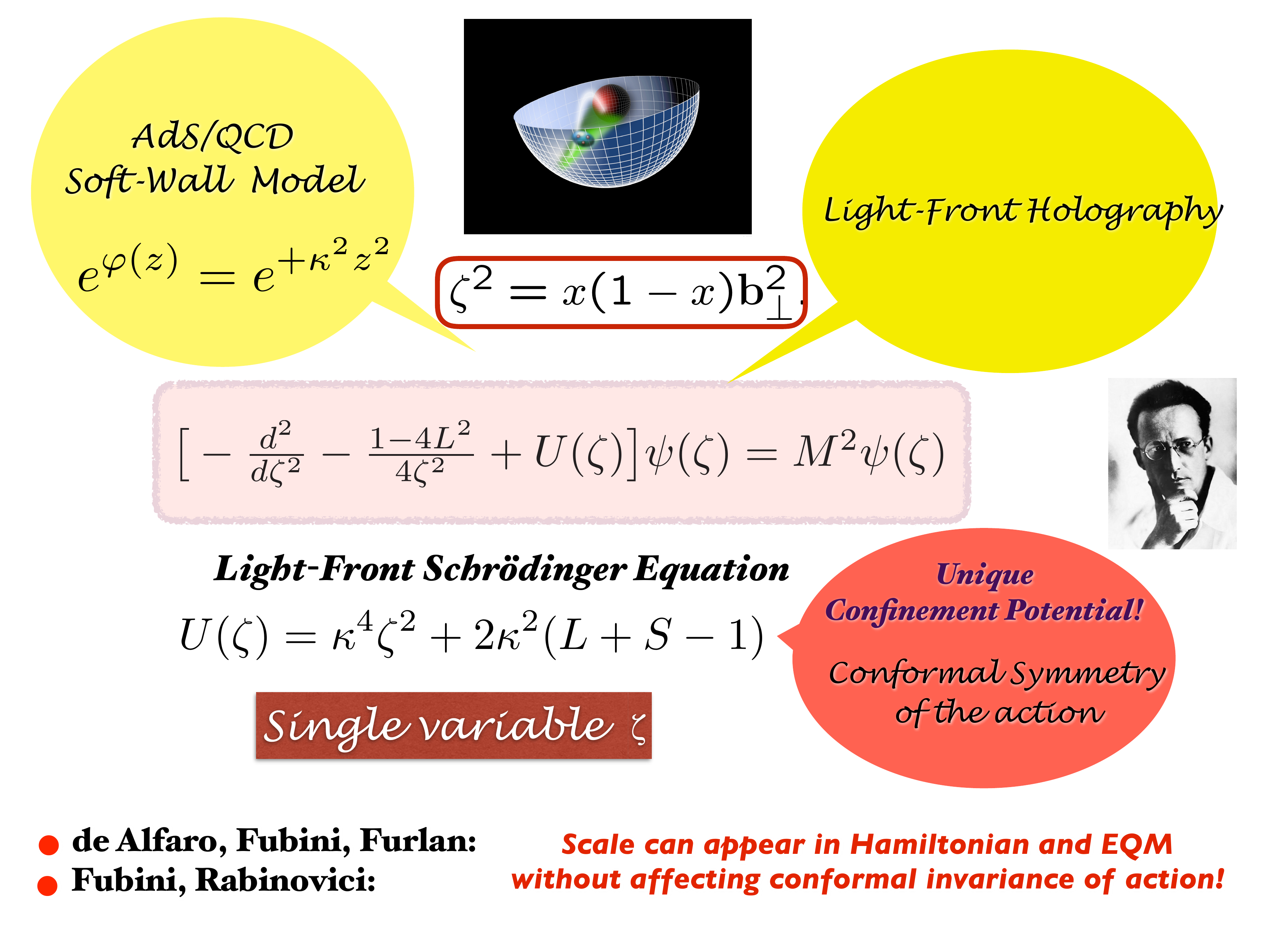}\hspace{2pc}%
\begin{minipage}[b]{18pc}\caption{\label{Fig1}The convergence of  theoretical methods for generating a model of hadron spectroscopy and dynamics with color confinement  and meson-baryon supersymmetric  relations.
\label{LFSE}}
\end{minipage}
\end{figure}

The LF equation for $q \bar q$ bound states for $m_q=0$ can be systematically reduced to a differential equation in a single LF radial variable $\zeta$:
$$[- {d\over d \zeta^2 }   + {(1-4 L^2) \over 4 \zeta^2}  +U(\zeta^2)]  \psi ]= M^2 \psi(\zeta)$$
where $\zeta^2 = b^2_\perp x(1-x)$ is the radial variable of the front form and $L = \max |L^z|$ is the LF orbital angular momentum~\cite{deTeramond:2008ht}.   
This {\it Light-Front Schr\"odinger equation}  is in analogy to the non-relativistic radial Schr\"odinger equation for bound states such as positronium in QCD.
See fig.~\ref{LFSE}.

Even more remarkably, the identical LF color-confining potential and the same LF equation of motion are obtained using Maldacena's anti-deSitter space in five dimensional AdS$_5$  Anti-DeSitter space when one identifies the fifth coordinate $z$ with the LF radial coordinate $\zeta$ 
and introduces a specific modification of the AdS$_5$ metric -- the ``dilaton"  $e^\phi(z) = e^{+ \kappa^2 z^2}$. 
The resulting prediction from AdS/QCD is the single variable LF Schrodinger Equation in $\zeta$, where 
$$U(\zeta^2) = \kappa^4 \zeta^2 + 2 \kappa^2 (L+ S -1)$$
and $\zeta^2=b^2_\perp x(1-x)$.  The holographic  identification of AdS$_5$  with the light-front Hamiltonian theory also automatically introduces 
the spin-dependent constant term $2 \kappa^2 (L+ S -1)$  in the LF Hamiltonian, where $L= \max  {L^z} , S= \max {S^z}$ with  $J^z= L^z+S^z$ are the LF spins. 
The application of dAFF thus leads to a color-confining LF potential $\kappa^4 \zeta^2 $, where
again the action remains conformal.

The eigenvalues for the meson spectrum are  $M^2(L, n) = 4\kappa^2 (n + {J+L\over 2} ).$  The mesonic spectrum of $q \bar q$ bound states is thus described as Regge trajectories in both the radial variable $n$ and the orbital angular momentum $L$ with the identical slope $4 \kappa^2$.   Color confinement is then a consequence of the light-front potential  $U(\zeta^2)$.  Again, only ratios of masses  and decay constants are predicted.

Remarkably, the pion $(n=0, J=L=0)$  is massless:  $m_\pi=0$ for $m_q=0$.    Thus light-front holography explains another fundamental question in hadron physics -- 
how a zero mass $q \bar q$ pseudoscalar bound state can emerge, despite the pion's composite structure.  
The eigensolutions generate both the mass spectrum and the light front wavefunctions 
$\psi_M(x, k_\perp, \lambda)$ for all $q \bar q$ meson bound state.

Nonzero quark masses appear in the ``LF kinetic energy" (LFKE) $\sum_i {k^2_\perp + m^2 \over x_i}$ contribution to the LF Hamiltonian -- the square of the invariant mass of the constituents: 
${\cal M}^2  = (\sum_i k^\mu_i)^2 $.  One can identify the the $m^2\over x$  contribution to the LFKE as arising in the Higgs theory from the coupling ${m\over x} \times m$ of each quark to the background zero-mode Higgs field~\cite{Srivastava:2002mw} which replaces the usual VEV of the standard time ``instant form".  In the heavy quark limit, one recovers the usual  $\sigma r$  
confining potential for heavy quarkonium~\cite{Trawinski:2014msa}.

 The correspondence of AdS$_5$ space with LF Hamiltonian theory in 3+1 dimensions and the identification of  the fifth dimensional AdS  coordinate $z$  with the LF radial coordinate $\zeta$ of the front form  in physical 3+1 spacetime is called ``light-front holography". Exclusive hadron amplitudes, such as elastic and transition form factors are given in terms of  convolutions of light-front wavefunctions~\cite{Brodsky:1980zm}.  In fact, The Drell-Yan-West formulae for electromagnetic and gravitational  form factors is identical to the Polchinski-Strassler~\cite{Polchinski:2001tt} formula for form factors in AdS$_5$.  This identification also provides a nonperturbative derivation of scaling laws~\cite{Brodsky:1973kr,Lepage:1980fj} for form factors at large momentum transfer.
Additional references and reviews of {\it Light-Front Holography}  may be found in 
refs.~\cite{deTeramond:2009xk,Brodsky:2010kn,deTeramond:2014yga,Brodsky:2014yha}.

\section{The QCD Running Coupling at all Scales} 

The form of the dilaton modifying AdS$_5$ also
leads~\cite{Brodsky:2010ur} to a Gaussian functional form for the nonperturbative QCD running coupling:  $\alpha_s(Q^2) \propto \exp{-{Q^2/4 \kappa^2}}$  in agreement with the effective charge  determined from measurements of the Bjorken sum rule.  Deur, de Teramond, and I~\cite{Brodsky:2010ur,Deur:2014qfa,Brodsky:2014jia} have also shown how the parameter $\kappa$,  which   determines the mass scale of hadrons and Regge slopes  in the zero quark mass limit, can be connected to the  mass scale $\Lambda_s$  controlling the evolution of the perturbative QCD coupling.  
The high momentum transfer dependence  of the coupling $\alpha_{g1}(Q^2)$ is  predicted  by  pQCD.  The 
matching of the high and low momentum transfer regimes  of $\alpha_{g1}(Q^2)$ -- both its value and its slope -- then determines a scale $Q_0 =0.87 \pm 0.08$ GeV which sets the interface between perturbative and nonperturbative hadron dynamics.  This connection can be done for any choice of renormalization scheme, such as the $\overline{MS}$ scheme,
The mass scale $\kappa$ underlying hadron masses  can thus be connected to the parameter   $\Lambda_{\overline {MS}}$ in the QCD running coupling by matching its predicted nonperturbative form to the perturbative QCD regime. The result is an effective coupling $\alpha_s(Q^2)$  defined at all momenta. 

One can measure the running QCD coupling constant $\alpha_s(Q^2) $ over a wide range of $Q^2$  from event shapes for electron-positron annihilation measured at a single annihilation energy $\sqrt s$~\cite{Wang:2019isi}.   The renormalization scale $Q^2$  of the running coupling depends dynamically on the virtuality of the underlying quark and gluon subprocess and thus the specific kinematics of each event~\cite{Gehrmann:2014uva}. The determination of the renormalization scale for event shape distributions can be obtained by 
using the Principle of Maximum Conformality (PMC)~\cite{Mojaza:2012mf,Brodsky:2011ig}, a rigorous scale-setting method for gauge theories, an all-orders extension of the BLM method~\cite{Brodsky:1982gc}, which satisfies all the requirements of Renormalization Group Invariance, including renormalization-scheme independence and consistency with Abelian theory in the $N_C \to 0$  limit~\cite{Wu:2019mky}.  In a recent paper~\cite{Wang:2019isi} we have applied the PMC to two classic event shapes measured in $e^+ e^-$  annihilation: the thrust (T) and C-parameter (C). 
The application of PMC scale-setting determines the running coupling to high precision over a wide range of $Q^2$, consistent with both
 LF  holography and pQCD~\cite{Deur:2017cvd}.

\begin{figure}[h]
\begin{minipage}{18pc}
\includegraphics[width=18pc]{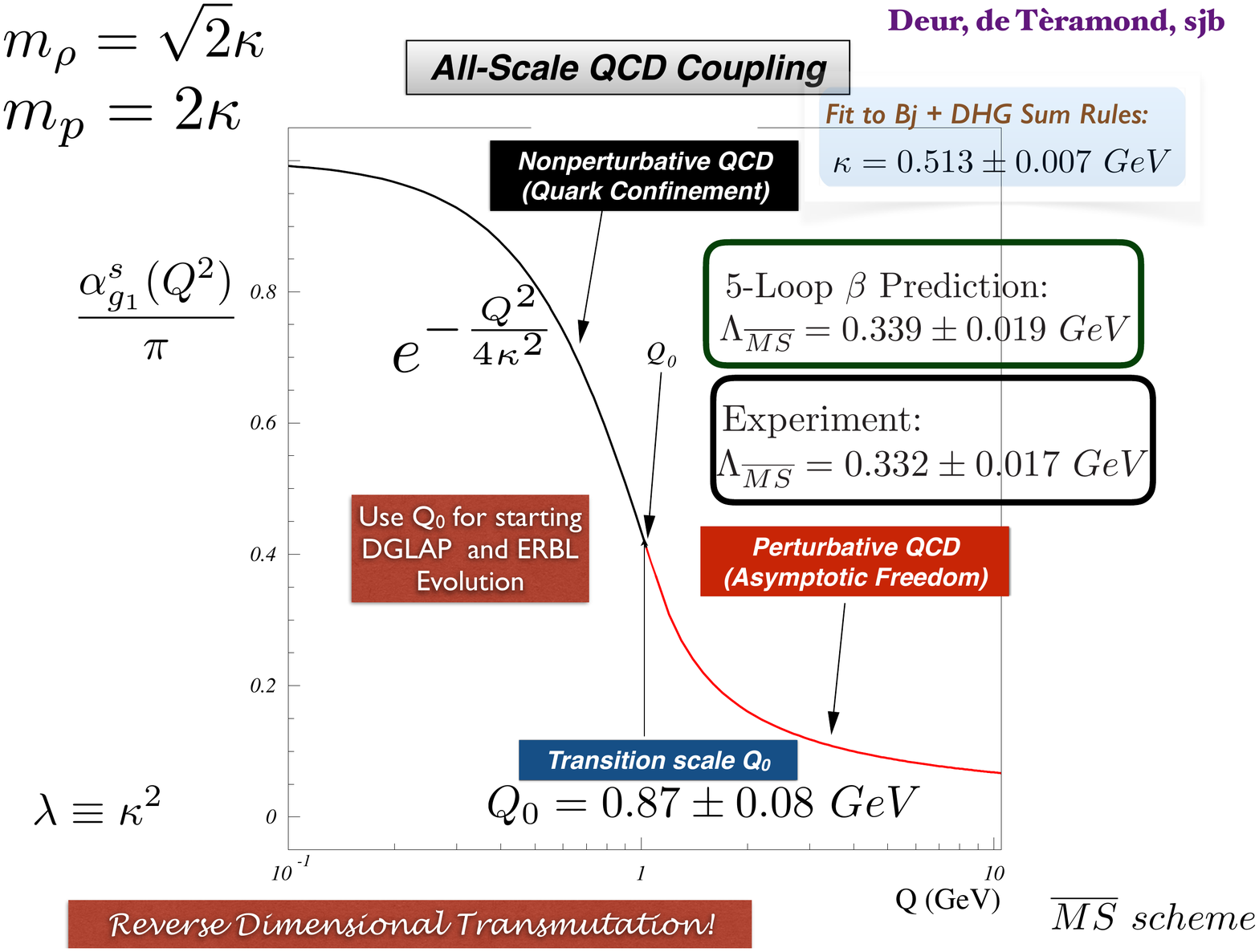}
\caption{Prediction from LF Holography and pQCD for the running coupling $\alpha_s^{g_1}(Q^2)$ at all scales.   
The magnitude and derivative of the perturbative and nonperturbative coupling are matched at the scale $Q_0$.  
This matching connects the perturbative scale 
$\Lambda_{\overline{MS}}$ to the non-perturbative scale $\kappa$ which underlies the hadron mass scale. 
From Ref.~\cite{Brodsky:2010ur}.
\label{DeurCoupling}}
\end{minipage}\hspace{2pc}%
\begin{minipage}{18pc}
\includegraphics[width=18pc]{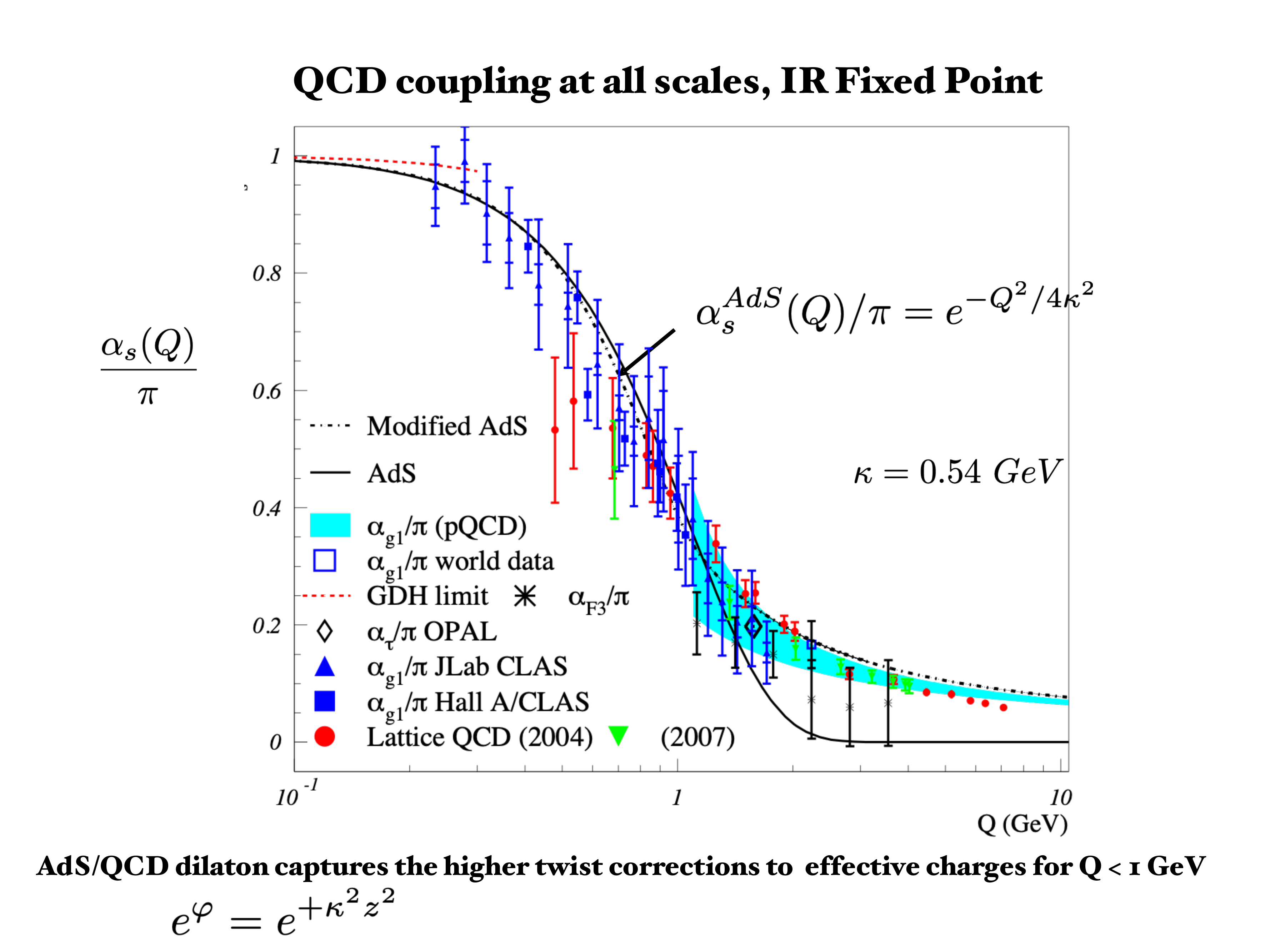}
\caption{Comparison of the matched nonperturbative QCD from LF holography and perturbative QCD prediction with experiment.}
{\label{CouplingData}}
\end{minipage} 
\end{figure}

\section{Superconformal Algebra and Supersymmetric Hadron Spectroscopy }

\begin{figure}[h]
\begin{minipage}{18pc}
\includegraphics[width=18pc]{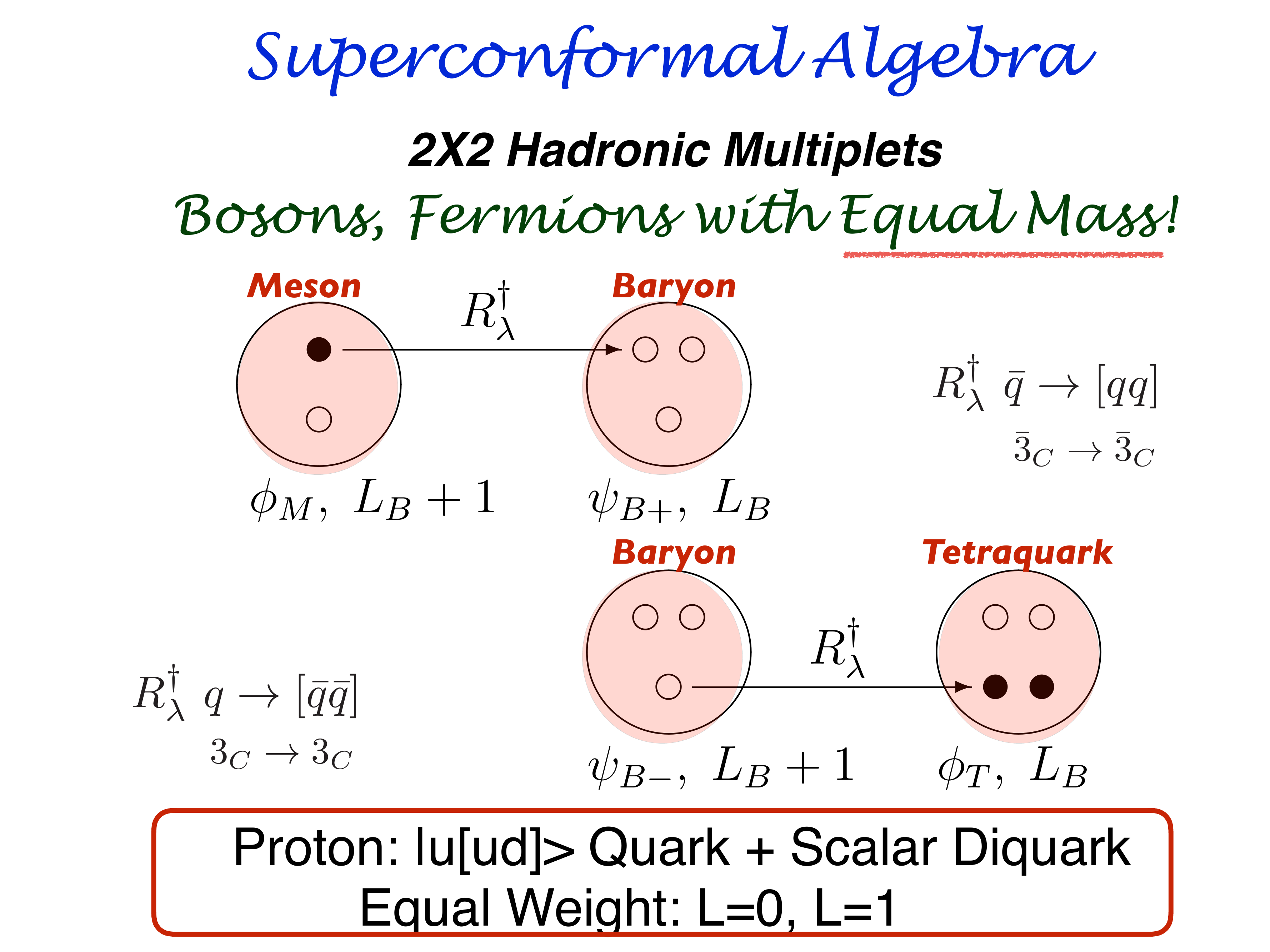}
\caption{\label{NSTARFigD}{The 4-plet representation of mass-degenerate hadronic states predicted by superconformal algebra~\cite{Brodsky:2013ar}.  Mesons are $ q \bar q$ bound states, baryons are quark -- antidiquark bound states and  tetraquarks are diquark-antidiquark bound states. The supersymmetric ladder operator $R^\dagger_\lambda $ connects quarks and  anti-diquark clusters of the same color.}}
\end{minipage}\hspace{2pc}%
\begin{minipage}{18pc}
\includegraphics[width=18pc]{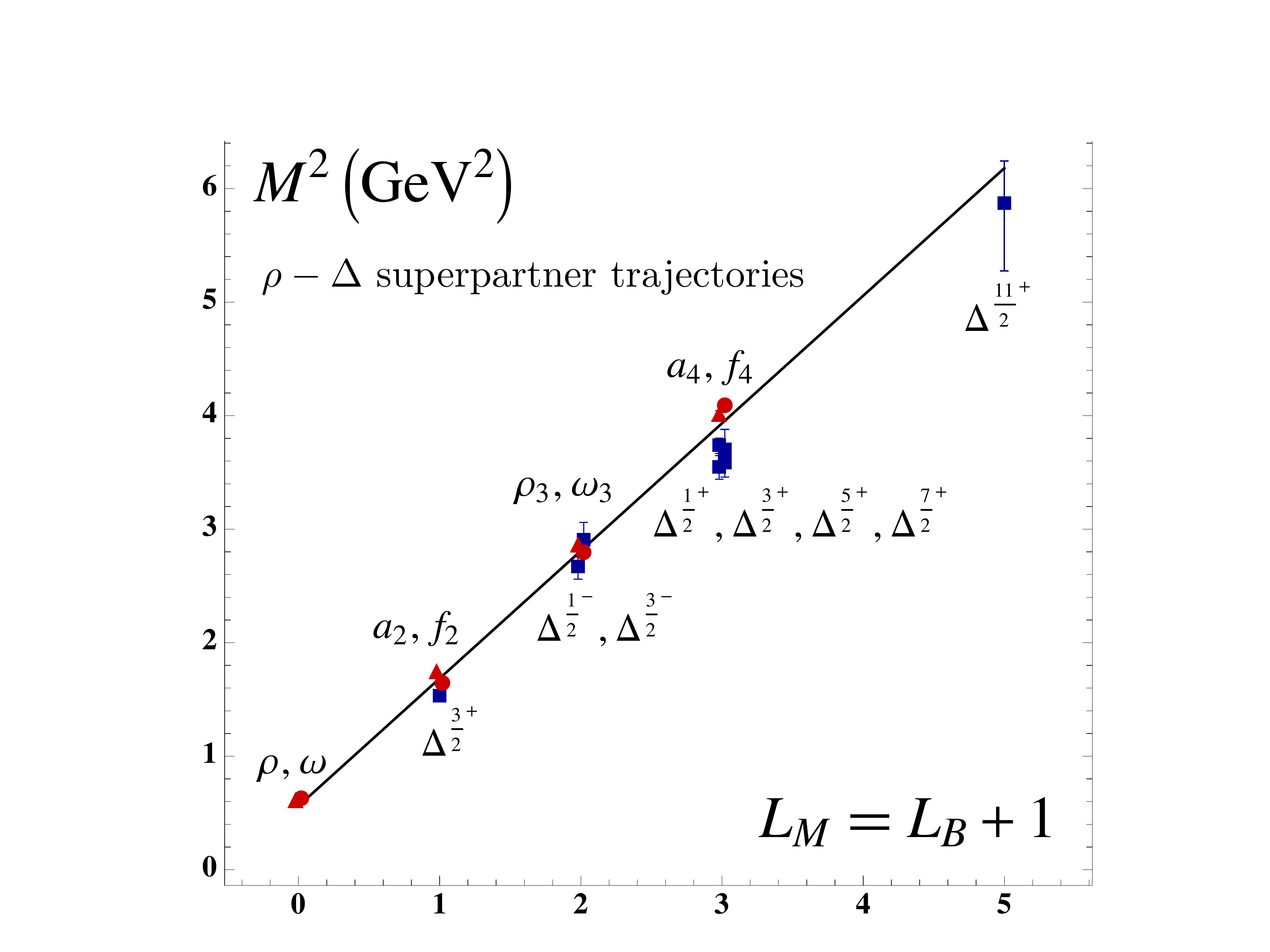}
\caption{\label{FigsJlabProcFig4.pdf}{Comparison of the $\rho/\omega$ meson Regge trajectory with the $J=3/2$ $\Delta$  baryon trajectory.   Superconformal algebra  predicts the degeneracy of the  meson and baryon trajectories if one identifies a meson with internal orbital angular momentum $L_M$ with its superpartner baryon with $L_M = L_B+1.$
See Refs.~\cite{deTeramond:2014asa,Dosch:2015nwa}.}}
\end{minipage} 
\end{figure}

Another advance in LF holography is the application~~\cite{deTeramond:2014asa,Dosch:2015nwa,Brodsky:2016rvj} of {\it superconformal algebra}, a feature of the underlying conformal symmetry of chiral QCD. 
The conformal group has an elegant $ 2\times 2$ Pauli matrix representation called {\it superconformal algebra}, 
originally discovered by  Haag, Lopuszanski, and Sohnius ~\cite{Haag:1974qh}.
The conformal Hamiltonian operator and the special conformal operators can be represented as anticommutators of Pauli matrices
 $H = {1/2}[Q, Q^\dagger]$ and  $K = {1/2}[S, S^\dagger]$.
As shown by Fubini and Rabinovici,~\cite{Fubini:1984hf},  a nonconformal Hamiltonian with a mass scale and universal confinement can then be obtained by shifting $Q \to Q +\omega K$, the analog of the dAFF procedure. 
In effect,  one has obtained generalized supercharges of the superconformal algebra~\cite{Fubini:1984hf}.
This ansatz extends the predictions for the hadron spectrum to a ``4-plet" -- consisting of mass-degenerate quark-antiquark mesons, quark-diquark baryons, and diquark-antidiquark tetraquarks, as shown in fig.~\ref{NSTARFigD}.   The 4-plet contains two entries $\Psi^\pm$  for each baryon, corresponding to internal orbital angular momentum $L$ and $L+1$.  This property of the baryon LFWFs is the analog of the eigensolution of the Dirac-Coulomb equation which has both an upper component $\Psi^+$ and a  lower component $\Psi^- =  {\vec \sigma \cdot \vec p  \over m+E -V} \Psi^+$.  

LF Schr\"odinger Equations for  both baryons and mesons can be derived from superconformal algebra~\cite{deTeramond:2014asa,Brodsky:2015oia,Dosch:2015nwa,Brodsky:2016rvj}.
The baryonic eigensolutions correspond to bound states of $3_C$ quarks to a $\bar 3_C$ spin-0 or spin-1 $qq$ diquark cluster;  the tetraquarks in the 4-plet are bound states of diquarks and  anti-diquarks.  
The quark-diquark baryons have two amplitudes $L_B, L_B+1$  with equal probability, a  feature of ``quark chirality invariance". 
The proton Fock state component $\psi^+$ (with parallel quark and baryon spins) and $\psi^-$ (with anti-parallel quark and baryon spins)  have equal Fock state probability -- a  feature of ``quark chirality invariance".  Thus the proton's spin is carried by quark orbital angular momentum in the nonperturbative domain. Predictions for the static properties of the nucleons are discussed in Ref.~\cite{Liu:2015jna}.  
The overlap of the $L=0$  and $ L=1 $  LF wavefunctions  in the Drell-Yan-West formula is required to have a  non-zero  Pauli form factor $F_2(Q^2)$ and anomalous 
magnetic moment~\cite{Brodsky:1980zm}.  The existence of both components is also necessary to generate the  pseudo-T-odd Sivers single-spin asymmetry in deep inelastic lepton-nucleon scattering~\cite{Brodsky:2002cx}.

The predicted spectrum, $M^2(n,L) = 4\kappa^2(n+L)$ for mesons and $M^2(n,L) = 4\kappa^2(n+L+1)$ for baryons, is remarkably consistent with observed hadronic spectroscopy.  
The Regge-slopes in $n$ and $L$ are identical.    
The predicted  meson, baryon and tetraquark masses  coincide if one identifies a meson with internal orbital angular momentum $L_M$ with its superpartner baryon or tetraquark with $L_B = L_M-1$. 
Superconformal algebra thus predicts that mesons with $L_M=L_B+1$ have the same mass as the baryons in the supermultiplet. 
An example of the mass degeneracy of the $\rho/\omega$ meson Regge trajectory with the $J=3/2$ $\Delta$-baryon trajectory is shown in  
Fig.~\ref{FigsJlabProcFig4.pdf}.   The value of $\kappa $ can be set by the $\rho$ mass;  however, only ratios of masses are predicted.

The combination of light-front holography with superconformal algebra thus leads to the novel prediction that hadron physics has supersymmetric properties in both spectroscopy and dynamics. The excitation spectra of relativistic light-quark meson, baryon and tetraquark bound states all lie on linear Regge
trajectories with identical slopes in the radial and orbital quantum numbers.  Detailed predictions for the tetraquark spectroscopy and  comparisons with the observed hadron spectrum are presented in ref.~\cite{Nielsen:2018uyn}.

\section{Supersymmetric Hadron Spectroscopy for Heavy Quarks }

The predictions from light-front holography and superconformal algebra have been extended to mesons, baryons, and tetraquarks with strange, charm and bottom quarks in refs.~\cite{Brodsky:2016rvj,Dosch:2016zdv,Nielsen:2018ytt}.  
Although conformal symmetry is strongly broken by the heavy quark mass, the basic
underlying supersymmetric mechanism, which transforms 
mesons to baryons (and
baryons to tetraquarks) into each other, still holds and gives remarkable connections and mass degeneracy across the entire spectrum of light, heavy-light and double-heavy hadrons.
The excitation spectra of the heavy quark meson, baryon and tetraquark bound states continue to lie on universal linear Regge
trajectories with identical slopes in the radial and orbital quantum numbers, but with an increased value for the slope.
For example, the mass of the lightest double-charm baryon $|c  [c q]>$,  where the $|[cq]$ is a scalar diquark, is predicted to be identical to the mass of the  $L=1$ orbital excitation  of the $|c \bar c>$ ( the $1^{++}$  $h_c^\prime(L=1)$ ) and also the   mass of the  $|[cq][\bar c \bar q]>$ double-charm tetraquark. 
In fact, the mass of the $h_c(3525)$ matches the mass of the double-charm baryon $\Xi^+_{ccd} (3520)$ identified by SELEX and a 
tetraquark candidate the $\Xi_{cc}(3415)$\cite{Brodsky:2017ntu}.
The  effective supersymmetric properties of QCD can be used to identify the structure of the heavy quark  mesons, baryons and tetraquark states~\cite{Nielsen:2018uyn}.

Thus one predicts supersymmetric hadron spectroscopy -- bosons and fermions with the same mass and twist- not only identical masses for the bosonic and fermionic hadron eigenvalues, but also supersymmetric relations between their eigenfunctions-- their light-front wavefunctions.  The baryonic eigensolutions correspond to bound states of $3_C$ quarks to a $\bar 3_C$ spin-0 or spin-1 $qq$ diquark cluster;  the tetraquarks in the 4-plet are bound states of diquarks and  anti-diquarks.

\section{Summary}

The  combination of light-front dynamics, its holographic mapping to AdS$_5$ space, and the dAFF procedure provide new insights,  not only into the physics underlying color confinement, but also the nonperturbative QCD coupling and the QCD mass scale. Reviews are given in  Refs.~\cite{Brodsky:2014yha,Brodsky:2015oia}.
The combination of light-front holography with superconformal algebra leads to the novel prediction that hadron physics has supersymmetric properties in both spectroscopy and dynamics.   The QCD Lagrangian is not supersymmetrical; however its hadronic eigensolutions conform to a fundamental 4-plet supersymmetric representation of superconformal algebra, reflecting the underlying conformal symmetry of semi-classical QCD for massless quarks.   The  resulting ``Light-Front Schr\"odinger equations" derived from LF holography incorporate color confinement and other spectroscopic and dynamical features of hadron physics, including a massless pion for zero quark mass and linear Regge trajectories with the {\it same slope}   in the radial quantum number $n$   and internal  orbital angular momentum $L$ for mesons, baryons, and tetraquarks. 

One can also observe features of superconformal symmetry in the spectroscopy and dynamics of heavy-light mesons and baryons.  
This approach predicts novel supersymmetric relations between mesons,  baryons, and tetraquarks  of the same parity as members of the same 4-plet representation of superconformal algebra.  One can test the similarities of their wavefunctions and form factors  in exclusive reactions such as  $e^+ e^- \to \pi T$ where $T$ is a tetraquark~\cite{Brodsky:2015wza}. 
Empirically viable predictions for spacelike and timelike hadronic  form factors, structure functions, distribution amplitudes, and transverse momentum distributions have also been obtained~\cite{Sufian:2016hwn}.  One can also observe features of superconformal symmetry in the spectroscopy and dynamics of heavy-light mesons and baryons.   
LF holography gives a remarkable first approximation to  hadron spectroscopy and dynamics, and  the hadronic LFWFs.  One also obtains viable predictions for tests of hadron dynamics such as spacelike and timelike hadronic form factors, structure functions, distribution amplitudes, and transverse momentum distributions. 
In recent papers, we have extended the LFWFs derived from LF Holography to incorporate non-valence higher Fock states and DGLAP evolution in order to compute other physics observables.  This includes detailed predictions for the spin structure of the valence quarks
in the proton~\cite{Liu:2019vsn} and its nonperturbative strange quark sea~\cite{Sufian:2018cpj}.   A new method for solving nonperturbative QCD ``Basis Light-Front Quantization" (BLFQ)~\cite{Vary:2014tqa},  uses the eigensolutions of a color-confining approximation to QCD (such as LF holography) as the basis functions,  rather than the plane-wave basis used in DLCQ, thus incorporating the full dynamics of QCD. 

We have also shown that  the Gribov-Glauber processes, which arise from leading-twist diffractive deep inelastic  scattering on nucleons 
and 
underly the shadowing and antishadowing of nuclear structure functions~\cite{Brodsky:1989qz}, prevent the application of the operator product expansion to the virtual Compton scattering amplitude $\gamma^* A \to\gamma^* A$  on nuclei and  thus negate the validity of the momentum sum rule for deep inelastic nuclear structure functions.~\cite{Brodsky:2019jla}

\section{Acknowledgements}

I thank  Cedric Lorc\'e for hosting LC2019 at Ecole Polytechnique. The physics results presented here are based on collaborations  and discussions with many collaborators, including 
Alexandre Deur, Guy de T\'eramond,  Hans G\"unter Dosch, Cedric Lorc\'e, Cheung Ji, Leonardo Di Giustino, Matin Mojaza, Kelly Chiu, Peter Lepage, Stan Glazek, Peter Lowdon, Philip Mannheim, Jerry Miller, Jennifer Rittenhouse West, Bo-Qiang Ma, Fred Goldhaber, Robert Shrock, Ivan Schmidt, Arkady Trawinski, Sheng-Quan Wang, Xing-Gang Wu, and Marina Nielsen. 

This research was supported by the Department of Energy,  contract DE--AC02--76SF00515.  
SLAC-PUB-17323.


\end{document}